\documentclass[aps,twocolumn,showpacs,preprintnumbers,superscriptaddress,amsmath,amssymb]{revtex4}
\usepackage{graphicx}
\usepackage{dcolumn}
\usepackage{color}
\usepackage{bm}

\begin{document}

\title{Photoionization of hydrogen atom by coherent intense high-frequency short laser pulses: Direct propagation of electron wave packets on large spatial grids}

\author{\firstname{Philipp~V.} \surname{Demekhin}}
\email{philipp.demekhin@pci.uni-heidelberg.de}
\affiliation{Theoretische Chemie, Physikalisch-Chemisches Institut, Universit\"{a}t  Heidelberg, Im Neuenheimer Feld 229, D-69120 Heidelberg, Germany}
\affiliation{Rostov State Transport University, Narodnogo Opolcheniya square 2,  Rostov-on-Don, 344038, Russia}

\author{\firstname{David} \surname{Hochstuhl}}
\email{hochstuhl@theo-physik.uni-kiel.de}
\affiliation{Institut f\"{u}r Theoretische Physik und Astrophysik, D-24098 Kiel, Germany}

\author{\firstname{Lorenz~S.} \surname{Cederbaum}}
\affiliation{Theoretische Chemie, Physikalisch-Chemisches Institut, Universit\"{a}t Heidelberg, Im Neuenheimer Feld 229, D-69120 Heidelberg, Germany}
\date{\today}

\begin{abstract}
The time-dependent Schr\"{o}dinger equation for the hydrogen atom and its interaction with coherent intense high-frequency short laser pulses is solved numerically exactly by propagating the single-electron wave packets. Thereby, the wavefunction is followed in space and time for times longer than the pulse duration. Results are explicitly shown for 3 and 10 fs pulses. Particular attention is paid to identifying the effect of dynamic interference of photoelectrons emitted with the same kinetic energy at different times during the rising and falling sides of the pulse predicted in   [\emph{Ph.V. Demekhin and L.S. Cederbaum}, Phys. Rev. Lett. \textbf{108}, 253001 (2012)]. In order to be able to see the dynamic interference pattern in the computed electron spectra, the photoelectron wave packet has to be propagated over long distances. Clearly, complex absorption potentials often employed to compute spectra of emitted particles cannot be used to detect dynamic interference. For the considered high-frequency pulses of 3 and 10 fs durations, this requires enormously large spatial grids. The presently computed photoionization and above-threshold ionization spectra are found to exhibit pronounced dynamic interference patterns. Where available,  the patterns are in very good agreement with previously published results on the photoionization  spectra which have been computed using a completely different method, thus supporting the previously made assumption that the above-threshold ionization processes are very weak for the considered pulse intensities and high carrier frequency. The quiver motion in space and time of a free electron in strong laser pulses is also investigated numerically. Finally, a discussion is presented of how fast the atom is ionized by an intense pulse.
\end{abstract}


\maketitle

\section{Introduction}
\label{sec:intro}

The presently available light sources, like attosecond lasers  \cite{KRAUSZ}, high-order harmonic generation sources \cite{highharm1,highharm2}, or free electron lasers \cite{FLASH,FERMI}, allow one to study the interaction of matter with super-intense coherent high-frequency ultra-short laser pulses. Many new phenomena, which are not available or difficult to observe with optical lasers operating in the nano- and picosecond regimes, arise in experiments with such pulses \cite{ZEWAIL,THbp1,THbp2,THbp3}. One of the main advantages here is that the high carrier frequencies allow one to directly access a few well-separated highly-excited electronic states of a system. This is usually not possible with optical pulses triggering simultaneously a dense spectrum of electronic states. Another advantage is that short pulses allow one to study the initiated dynamics on the relaxation timescale of highly-excited states (typically femto- or even attoseconds). The theoretical description of the processes triggered by realistic pulses requires the propagation of electron wave packets created by the pulse in real time and real.

One of the phenomena arising due to the interaction of matter with coherent intense high-frequency short  pulses is known as dynamic interference \cite{DynIntLETT}. There, the pulse ionizes a system by the absorption of a single photon and, at the same time, induces a time-dependent shift of the `dressed' ground state of an atom relative to the continuum due to an AC Stark effect in the electronic continuum. The energy shift adiabatically follows the pulse intensity envelope $g^2(t)$   \cite{Sussman11}. Because of this time-dependent energy shift, the photoelectrons emitted when the pulse rises have the same kinetic energy as those emitted when the pulse decreases. These two electron wave packets emitted at different times superimpose and interfere, and the resulting photoelectron spectrum exhibits a pronounced multipeak pattern \cite{DynIntLETT}. Similar dynamic interference of electrons was also found in the  above-threshold ionization (ATI) spectra of model anions \cite{NOTOURS1,NOTOURS2}, in the sequential multiphoton ionization of atoms exposed to optical \cite{jones} and high-frequency \cite{DemekhinDIres} pulses, as well as in the resonant Auger effect of atoms in free electron laser pulses \cite{DemekhinDIres}.

In order to solve the time-dependent Schr\"{o}dinger equation for the hydrogen atom interacting with a pulse we have used in our recent work \cite{DynIntLETT} a previously developed theoretical approach \cite{Chiang10,Demekhin11SFatom,MolRaSfPRL,DemekhinCO,DemekhinICD,DemekhinBLOCK}. In this approach the total wave function is expanded in terms of the full set of the field-free stationary states of the system. This leads to equations for the corresponding population amplitudes which were then propagated in a large but restricted relevant subset of these states.  In particular, possible transitions between different electronic continuum states, which are responsible for the ATI processes and contribute also to the
ponderomotive energy of an electron in the field, were assumed to be weak and therefore neglected in \cite{DynIntLETT}. In order to verify this assumption and to have a complete account of ponderomotive forces, we solve in the present work the time-dependent Schr\"{o}dinger equation for the same problem numerically exactly. For this purpose we directly propagate the single-electron wave packet during and briefly after the pulse and extract from this wave packet the full spectrum of the emitted electrons. This is a challenging problem by itself (see below). To allow for the dynamic interference to take place, one needs to propagate the full photoelectron wave packet in space and time without using complex absorption potentials at the boundaries. For high-frequency pulses of even a few femtosecond duration the propagation requires spatial grids of $\sim 10^4$ Bohr. We have found that cutting off even a small tail of the wave packet falsifies the results seemingly.

To the best of our knowledge, no such explicit time-dependent calculations on the electron wave packet propagation are reported in the literature for smooth realistic pulses, even not for a single-electron system. There is a series of works  \cite{POPOV1,POPOV2,POPOV3}  (see also references therein
for earlier works by the same authors), reporting calculations performed for a smoothed Coulomb potential and short trapezoid-shaped pulse envelopes using restricted spatial grids and a complex absorption potential at the boundary. The pulses used consist of a rapidly rising edge, a long plateau, and a rapidly falling edge ``to let the major fraction of photoelectrons to be formed at the pulse plateau''. We note that such a pulse cannot give rise to dynamic interference by definition even if complex absorbing potentials were not used. The calculations performed for  $\omega=30$~eV   \cite{POPOV2,POPOV3} showed that the photoelectron peak position is shifting to higher energies as the intensity of the field grows. This finding already indicates that the binding energy of the system decreases with the field strengths \cite{POPOV2}, which is, to our opinion, due to the AC Stark effect in the electronic continuum.

In order to solve the technically challenging problem at hand one needs sophisticated techniques. The required numerical algorithms are, however, already well established. It will be demonstrated here that one is able at present to accurately describe on very large spatial grids the interaction of one electron with intense pulses, where the absorption of multiple photons makes the problem already quite complicated. This gives hope that the problem of interaction of strong pulses with more particles can also be accurately solved in the future. There are, for instance, works reporting the efficient propagation of up to three electrons (nine degrees of freedom) interacting with weak fields \cite{C3el1,C3el2}  by the time-dependent close-coupling (TDCC) method \cite{C2el}. There are also weak field calculations for two electrons on relatively large (up to $\sim 10^3$ Bohr) radial grids \cite{Clgr}, but these are still 10 times smaller than the grid sizes needed  here in the case of intense fields. Alternatively, the multi-configuration time-dependent Hartree (MCTDH) method \cite{MCTDH,MCTDH1} could be used which is known to be an optimal approach for wave packet propagation in many degrees of freedom.   Clearly, this approach requires efficient propagation algorithms for the underlying single-particle time-dependent functions \cite{MCTDHBOOK}. Being formulated for bosonic particles the method is known as MCTDHB \cite{BOSON1}, and for fermions as MCTDHF \cite{FERM1,FERM2,FERM3,FERM4,DAVID1}. In the present work which is on a single electron we utilize a particular code implemented for MCTDHF \cite{DAVID1} which is based on the general formulation of the problem given in \cite{BOSON2} and has been slightly modified  here to efficiently propagate single-particle wave packets in the presence of strong pulses. This code has also the potential to attack the problem of propagation of several particles on large spatial grids.

\section{Theory}
\label{sec:the}

The total Hamiltonian for the hydrogen atom interacting with the linearly polarized laser field is given by (atomic units are used throughout)
\begin{equation}
\label{eq:totH}
\hat{H}(t)=\frac{\mathbf{\hat{p}}^2}{2}-\frac{1}{r}+\hat{z}\,\mathcal{E}(t),
\end{equation}
where the electric field
\begin{equation}
\label{eq:field}
\mathcal{E}(t)=\mathcal{E}_0g(t)\cos \omega t.
\end{equation}
is polarized along the z-axis. Here,  $\mathcal{E}_0$ is the field amplitude and  $\omega$ is the carrier frequency of the pulse with a time-envelope $g(t)$.  In order to solve the time-dependent Schr\"{o}dinger equation for the Hamiltonian (\ref{eq:totH}) we use the numerical approach implemented in the code  \cite{DAVID1} for efficiently propagating single-electron orbitals. A few points of the approach, which are of essential relevance to the present work, are outlined below.

First, we make use of the axial symmetry along the $z$  axis of the problem and employ a partial wave expansion of the single-electron
wave function $\Psi(\textbf{r},t)$ in terms of spherical harmonics $Y_{\ell 0}$
\begin{equation}
\label{eq:PWE}
\Psi(\textbf{r},t)=\sum_{\ell}  \frac{P_\ell(r,t)}{r} Y_{\ell 0}(\theta).
\end{equation}
By substituting the wave function (\ref{eq:PWE})  into the time-dependent Schr\"{o}dinger equation for the Hamiltonian (\ref{eq:totH}), one straightforwardly obtains the following system of coupled equations for the radial harmonics $ P_\ell(r,t) $
\begin{multline}
\label{eq:CDE}
i \frac{ \partial{P}_\ell(r,t)}{\partial t}= \left\{ -\frac{1}{2} \frac{\partial^2 }{\partial r^2} -\frac{1}{r} + \frac{\ell(\ell+1)}{2r^2}\right\}     {P}_\ell(r,t) \\
 +r\mathcal{E}_0g(t)\cos \omega t \left[\begin{array}{c}\sqrt {\frac{(\ell+1)^2}{(2\ell+3)(2\ell+1)}}\end{array} {P}_{\ell+1}(r,t) \right.  \\ \left. +\begin{array}{c}\sqrt {\frac{\ell^2}{(2\ell+1)(2\ell-1)}}\end{array}{P}_{\ell-1}(r,t)\right].
\end{multline}

The radial coordinate in the above system of coupled one-dimensional equations (\ref{eq:CDE}) can  further be represented by a discrete variable representation (DVR) basis set $\chi_i(r)$
\begin{equation}
\label{eq:DVREXP}
P_\ell(r,t)=\sum_i b_{\ell , i}(t) \chi_i(r).
\end{equation}
As usual, employing  DVRs has the advantage that spatially local operators posses a diagonal representation where a function acquires the value on the respective DVR grid point   \cite{PR00}. In order to avoid full matrices of the kinetic energy operator, we use here the finite element discrete variable representation (FEDVR) introduced in  \cite{FEDVR}. We thus divide the radial coordinate space in a chosen number of finite elements. In each finite element, the basis functions  $\chi_i(r)$ are represented by the normalized Legendre interpolating polynomials
\begin{equation}
\label{eq:basis}
\chi_i(r)=\frac{1}{\sqrt{w_i}}\prod_{j\ne i}\frac{r-r_j}{r_i-r_j},
\end{equation}
constructed over a Gauss-Lobatto grid $\left\{ r_i\right\}$  with weights $\left\{w_i\right\}$. Note that each basis function vanishes at all grid points except one. Correspondingly, one arrives at a banded  structure of the kinetic energy matrix, which makes numerical solution of
Eqs.~(\ref{eq:CDE}) faster, and, at the same time, we can use sufficiently flexible basis sets adjustable to different physical problems.

As a consequence of the DVR representation (\ref{eq:DVREXP}), the system of equations (\ref{eq:CDE}) can be straightforwardly transformed to a system of equations for the evolution of the time-dependent expansion coefficients $b_{\ell , i}(t)$ with known one-particle integrals. This system was propagated using the short-iterative Lanczos method employing algorithm \cite{ALG1}  to approximate the exponential time-evolution operator. The initial ground state of the H atom in the absence of the pulse (although analytically known) is consistently obtained via imaginary time propagation starting from a guess function, which requires a negligible time as compared to the real time propagation.

The final momentum distribution of the emitted photoelectrons was computed by Fourier transforming the electron wave packet $\Psi(\textbf{r})=\Psi(\textbf{r},t= \infty)$ after the pulse has expired
\begin{equation}
\label{eq:FUR}
\Psi(\textbf{k})=  \int \Psi(\textbf{r}) \,e^{-i\mathbf{k}\mathbf{r}} d^3\mathbf{r}.
\end{equation}
In order to exclude contributions to the final wave packet $\Psi(\textbf{r}) $ from electrons remaining bound to the nucleus (i.e., from the ground and Rydberg states), the inner-part region of the radial spatial variable $r$ was excluded from the transformation (\ref{eq:FUR}). Finally, the kinetic energy spectrum $\sigma(\varepsilon)$  of the emitted photoelectrons is obtained by angular-averaging the momentum density distribution as
\begin{equation}
\label{eq:SPE}
\sigma(\varepsilon)= k\int \vert\Psi(\mathbf{k})\vert^2 d\Omega_k
\end{equation}
and using that $k=\sqrt{2\varepsilon}$.

As in our previous study \cite{DynIntLETT}, the present calculations were performed for a Gaussian-shaped pulses with a time-envelope $g(t)=e^{-t^2/\tau^2}$ and central frequency $\omega = 53.6057$~eV, which is 40~eV above the ionization threshold of H. The electron wave packets were propagated in the time interval of $[-3\tau,+3\tau]$, for which the field amplitude at the interval boundaries is almost four orders of magnitude weaker than at the pulse maximum ($t=0$). Calculations were performed for two pulse durations of $\tau=3$~fs and  $\tau=10$~fs. The shorter pulses were, but the longer  pulses were not studied in our previous work  \cite{DynIntLETT}.

As will become evident below, the computed electron spectra exhibit a clear sequence of ATI peaks located around the multiphoton absorption energies $\varepsilon_0^n = n\cdot\omega -IP$. For the largest field intensity considered here, the third ATI peak located at $\varepsilon_0^4 =4\omega -IP$ is about five orders of magnitude weaker than the main photoionization peak located at $\varepsilon_0^1 =\omega -IP$  (see discussion around Fig.~\ref{fig_ati3fs}). Because of this fact, we have decided to choose the parameters of the present calculations such that the photoionization peak and the three subsequent ATI peaks in the final electron energy spectra are described accurately. For this purpose, we had to include the  $\ell\le 4$ harmonics in the partial wave expansion (\ref{eq:PWE}).

For the photon energy $\omega$ used, the third ATI peak is formed by electrons with momentum $k\sim 3.85$~a.u. Consequently, we have chosen the size of the radial grid such that photoelectrons with $k<4$~a.u.  do not hit the outward grid boundary during the whole propagation time. During the $\tau=3$~fs pulse, the photoelectrons with momentum $k\sim 4$~a.u. may move off the nucleus by $R_{max}=6\tau k\sim 3000$~a.u., which was chosen as the radial grid size. The latter was represented by 1000 equidistant finite elements  of size 3~a.u. Each finite element was covered by 15 Gauss-Labbato points. For the $\tau=10$~fs pulse, the radial grid parameters were: $R_{max}=6\tau k\sim 10000$~a.u.   represented by 3125 finite elements with the size of 3.2 a.u., and each finite element is covered by 16 Gauss-Labbato points. The convergence of the solution with respect to the spatial grid representation and integration time step has been ensured.

\section{Results and discussion}
\label{sec:res}

\subsection{Hydrogen atom in short high-frequency pulses}
\label{sec:resHYDR}

\begin{figure}
\includegraphics[scale=0.4]{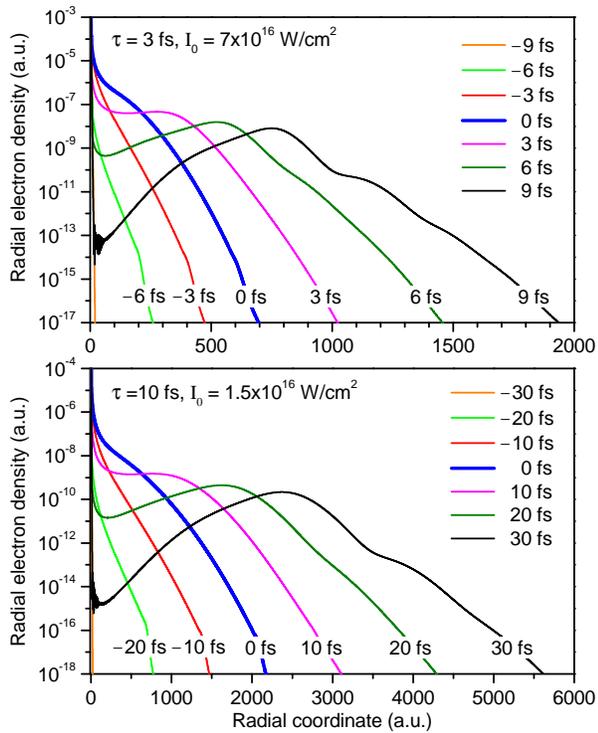}
\caption{(Color online) Time-evolution of the radial electron density (total electron wave packet, $\vert \Psi(r,t) \vert^2$) computed for the hydrogen atom exposed to Gaussian-shaped short pulses of carrier frequency of $\omega=53.6057$~eV. \emph{Upper panel}: The duration of the pulse is  $\tau=3$~fs, and the peak intensity $I_0=7\times10^{16}$~W/cm$^2$. The wave packet shown for $t=-9$~fs  (straight orange line near the coordinates origin) is the ground state wave packet. Note that the final wave packet computed at $t=9$~fs spreads out to about $r\sim 3000$~a.u. \emph{Lower panel}: The duration of the pulse is $\tau=10$~fs, and the peak intensity $I_0=1.5\times10^{16}$~W/cm$^2$. The wave packet shown for $t=-30$~fs   is the ground state wave packet. Note that the final wave packet computed at $t=30$~fs  spreads out to about $r\sim 10000$~a.u.}\label{fig_WP3fs}
\end{figure}

Figure~\ref{fig_WP3fs} illustrates the time evolution of the total electron wave packet computed for the hydrogen atom exposed to short pulses of carrier frequency  $\omega=53.6057$~eV. The upper panel shows the results for the  $\tau=3$~fs pulse and peak intensity of $I_0=7\times10^{16}$~W/cm$^2$. Before the pulse arrives, the electron is in the 1s ground state of H which is very close to the nucleus (note the scale of the figure). At very early times, when the pulse arrives, the part of the wave packet which will contribute to the photoelectron spectrum is created around the nucleus and starts to propagate outwards. Before the maximum of the pulse has arrived (time 0~fs corresponds to the pulse maximum, thick blue curve), the wave packet being continuously pumped by the pulse does not posses a maximum. A maximum in the radial electron density distribution starts to develop only after the pulse maximum has arrived. For the wave packet computed at $t=3$~fs the maximum of the wave packet is around $r=275$~a.u.

As the pulse expires, the maximum becomes more and more pronounced and it moves further outwards. At $t=6$~fs this maximum is around  $r=525$~a.u. In the final wave packet computed at $t=3\tau=9$~fs, the maximum is located at about $r=750$~a.u. This maximum is mainly responsible for the photoionization peak at $\varepsilon_0^1=\omega-IP$ in the final energy spectrum computed via Eqs.~(\ref{eq:FUR}) and (\ref{eq:SPE}). The weak humps seen in the final wave packet at about  $r=1150$~a.u. and 1500~a.u. are mainly formed by the photoelectrons contributing to the first and the second ATI peaks in the final energy spectrum. The lower panel of Fig.~\ref{fig_WP3fs}  depicts the time evolution of the electron wave packet for the $\tau=10$~fs pulse and peak intensity of $I_0=1.5\times10^{16}$~W/cm$^2$. The results are rather analogous to those described above for the $\tau=3$~fs pulse except that the time and the spatial scales are different. The detached electron now reaches much longer distances and the calculation has become even much more cumbersome.

\begin{figure}
\includegraphics[scale=0.4]{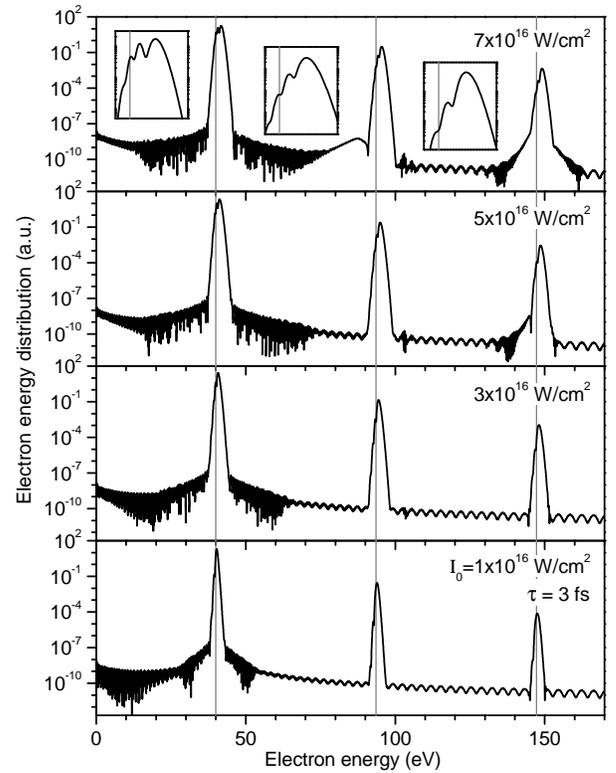}
\caption{ Computed energy distributions of the emitted photoelectrons after exposure of the H atom to coherent laser pulses of   $\tau=3$~fs duration, $\omega=53.6057$~eV carrier  frequency, and different peak intensities (indicated in each panel). Shown are the photoionization peak and the two subsequent ATI peaks produced by multiphoton absorption from the ground state (the next ATI peaks are weak and are not shown). The field-free multiphoton absorption energy positions $\varepsilon_0^n = n\cdot\omega -IP$ are indicated by vertical gray lines. The insets in the uppermost panel illustrate on an enlarged scale the similarity between the dynamic interference patterns in the photoionization and ATI spectra (see also text for details).}\label{fig_ati3fs}
\end{figure}

The energy distribution of the electrons emitted during the illumination of the hydrogen atom by coherent  $\tau=3$~fs laser pulses of different peak  intensities are depicted in Fig.~\ref{fig_ati3fs}. For transparency of the figure, the energy range is restricted to show the photoionization peak and the two subsequent ATI peaks (note the logarithmic scale on the vertical axis). One can see that, as the field strength increases from the bottom to the top of the figure, the photoionization peak and both ATI peaks are systematically shifted to higher electron energies. For reference, the field-free multiphoton absorption energy positions $\varepsilon_0^n = n\cdot\omega -IP$ are indicated by vertical lines.

All peaks in the electron energy distribution exhibit pronounced multiple-peak structures which are due to the dynamic interference \cite{DynIntLETT}. This fact is demonstrated in the insets in the uppermost panel for the largest intensity considered here. The left inset shows the photoionization
peak on an enlarged scale, whereas the middle and right insets enlarge the first and the second ATI peaks, respectively. As can be seen from these insets, the dynamic interference patterns are qualitatively similar for the photoionization peak and for the two subsequent ATI peaks, each possessing three well-resolved oscillations of the intensity (the vertical axis in all insets is in logarithmic scale). The maximum of the intensity of each
of these peaks is, however, very different: for the photoionization peak it amounts to 16.9~a.u.; for the first, second, and third ATI peaks (the latter is not shown in the figure), the maximum amounts to 0.31, 0.0046, and 0.00013 a.u., respectively. Clearly, the ATI peaks are weak for the short pulses of the considered high frequency and intensities.

\begin{figure}
\includegraphics[scale=0.4]{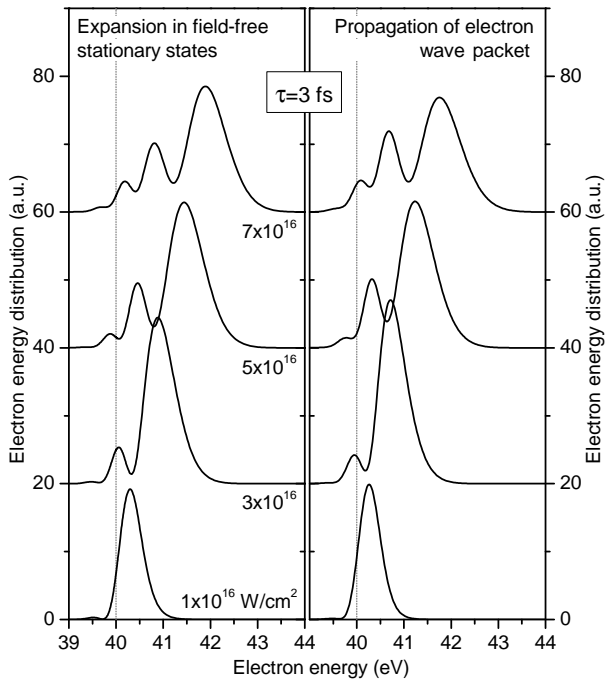}
\caption{Photoionization spectra of the hydrogen atom exposed to Gaussian-shaped pulses of  $\tau=3$~fs duration, carrier frequency of $\omega=53.6057$~eV, and different peak intensities (indicated near each spectrum) computed by two different theoretical approaches.\emph{ Left panel}: by expanding the total wave function in the field-free stationary states (results from Ref.~\cite{DynIntLETT}). \emph{Right panel:} by direct propagation of electron wave packet (present results, first peaks shown in Fig.~\ref{fig_ati3fs}).}\label{fig_PES3fs}
\end{figure}

In Fig.~\ref{fig_PES3fs}, we compare the photoionization spectra computed in the present work by direct propagation of the electron wave packets with those published in our previous work \cite{DynIntLETT} for the same $\tau=3$~fs pulses. One can see that the presently computed spectra (right panel) are in a very good agreement with the spectra computed by a totally different theoretical approach (left panel). In the latter approach the amplitudes of the populations of a restricted relevant subset of stationary states of the field-free system are propagated (see also above). Indeed, for every peak intensity indicated in the figure, each pair of  computed photoionization spectra possesses an equal number of oscillations caused by the dynamic interference, and the energy positions and relative heights of the multiple peak structures are very similar.

\begin{figure}
\includegraphics[scale=0.4]{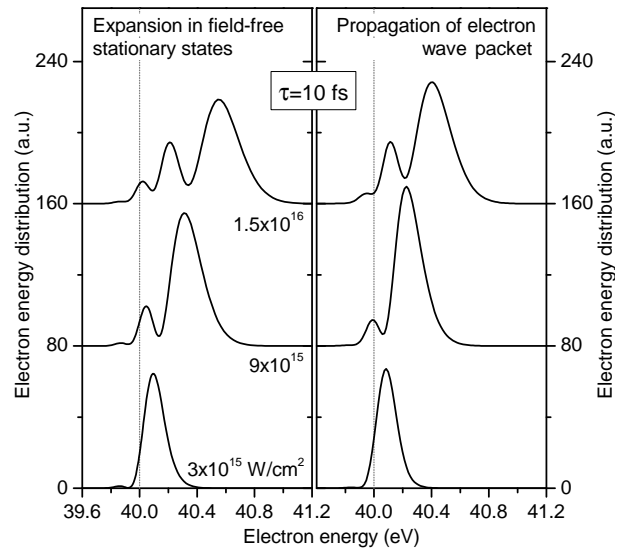}
\caption{The same as in Fig.~\ref{fig_PES3fs} but for  $\tau=10$~fs  pulses. The spectra shown in the left panel are not available in Ref.~\cite{DynIntLETT}, but are computed here by the method of this reference.}\label{fig_PES10fs}
\end{figure}

As shown above and discussed in the preceding section, the spatial grid required to directly propagate wave packets created by the longer $\tau=10$~fs pulses is much larger than for the $\tau=3$~fs pulses. Although the propagation time required is also much longer ($6\tau=60$~fs) for the longer pulse, the calculations are still feasible by the method and code of Ref.~\cite{DAVID1}. In Fig.~\ref{fig_PES10fs} we compare the photoionization spectra of H computed for $\tau=10$~fs pulses at different peak intensities by the two different approaches, i.e., by the present approach (right panel) and by the
method of our previous work  \cite{DynIntLETT} (left panel). The results of the latter calculations are also new and not available in Ref.~\cite{DynIntLETT}. From Fig.~\ref{fig_PES10fs} one can see that the numerically exact results of the direct wave packet propagation again agree very well with the spectra computed by our previous approach \cite{DynIntLETT}.

It is well known \cite{B1,B2} that in strong laser fields each electronic state experiences an AC Stark shift, including the ground state, Rydberg levels, and electron continuous states. Since an electron in the continuum interacting with an oscillating field cannot have an energy smaller than the average energy of its quiver-motion (known as ponderomotive energy), one argues that a strong field shifts the ionization threshold by this ponderomotive energy. There is a common belief \cite{B1} that in the high-frequency limit, when the carrier frequency is much larger than the field-free ionization potential of a system  ($\omega \gg IP$), each electronic state acquires the same AC Stark shift given by the universal formula for the ponderomotive potential $U_p=\mathcal{E}^2_0/4\omega^2$ (all quantities in atomic units). For this reason, at all intensities during the pulse the AC Stark shifts of the ground state and of the ionization threshold are expected to compensate each other \cite{B1}. In turn, this would of course eliminate the dynamic interference effects as these vanish if the shift in the spectrum vanishes  \cite{DynIntLETT}.

The results obtained and the very good agreement between the spectra in Figs.~\ref{fig_PES3fs} and \ref{fig_PES10fs} computed by two very different approaches demonstrate that the above qualitative argumentation based on simple general assumptions is not correct. The present numerically exact calculations clearly show that the AC Stark shift of the ground state is not compensated by the ponderomotive shift of the ionization threshold. Such a compensation is probably possible only for very high carrier frequencies at which both shifts vanish anyway, since $U_p \xrightarrow{~\omega \to \infty~} 0$. It is hence clear now that if the parameters of the pulse, i.e., shape, intensity and frequency, are such that the total photoelectron spectrum experiences a shift, this shift results from the fact that these two competing effects do not compensate each other, and then the dynamic interference predicted in \cite{DynIntLETT}  takes place.

In spite of the very good agreement, a slight quantitative disagreement between the results of the two sets of calculations is, however, evident from
Figs.~\ref{fig_PES3fs} and \ref{fig_PES10fs} and needs to be discussed here. It is due to the incompleteness of the basis set of stationary field-free states used to expand the total wave function in  \cite{DynIntLETT}. This expansion was restricted to the ground state, Rydberg states  $n\ell$, and photoionization $\varepsilon p$ continuum states. The convergence of the solution with respect to the included Rydberg states of different $n$ and $\ell$  has been insured there. However, transitions between continuum states $\varepsilon \ell$, which are responsible for the formation of
ATI peaks in the final energy spectrum of the emitted electrons, were neglected in our previous calculations  \cite{DynIntLETT}. These transitions are naturally included in the present numerically exact calculations. We would like to remind that even for the largest field intensity considered here, already the first ATI peak is almost two orders of magnitude weaker than the main photoionization peak. These weak ATI processes, neglected in \cite{DynIntLETT}, are responsible for the slight disagreement between the results of the two sets of calculations.

\subsection{ Free electron wave packet in short high-frequency pulses}
\label{sec:resFREE}

In order to visualize how the free electron wave packet expands in short high-frequency pulses and to have a stringent check of the quality of the present numerical calculations, let us now investigate the motion of a free electron in  strong  laser pulses. In the high-frequency limit \cite{B1}, it is expected that a free electron gains the ponderomotive energy $U_p$, which is an average energy of its  quiver-motion induced by a strong oscillating external field. In order to check this prediction in the case of short pulses, we start with the electronic wave packet given by the 1s ground state of hydrogen, and propagate this wave packet without the attractive Coulomb potential  exerted by the nucleus. Without the pulse, this wave packet expands in $r$-space as time proceeds, but it remains unchanged in $k$-space. In the presence of a pulse, the momentum distribution of the electron wave packet changes as well, and it stabilizes as the pulse expires.  As a result, the kinetic energy of the wave packet changes during the pulse. In addition to that, in the presence of a pulse the electron wave packet acquires also potential energy due to its interaction with the field, which is a time-dependent quantity too.

\begin{figure}
\includegraphics[scale=0.31]{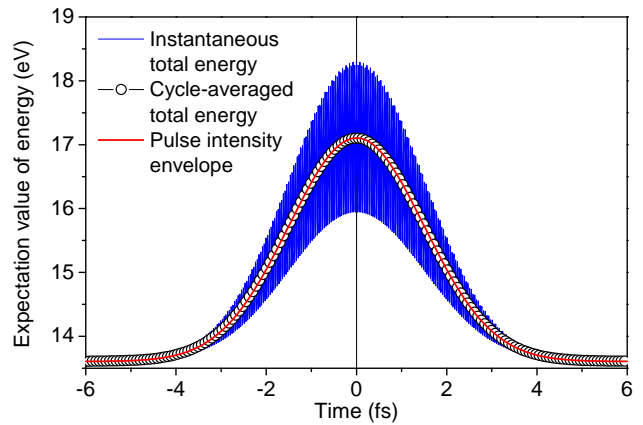}
\caption{ (Color online) Time-evolutions of the expectation values of the energy of a free electron wave packet. Initially, before the pulse arrived, the electron was in a wave packet like that of the ground state of the hydrogen atom.  The electron is then exposed to a  $\tau=3$~fs pulse of frequency $\omega=53.6057$~eV and peak intensity of $I_0=7\times 10^{16}$~W/cm$^2$. Blue thin solid curve: instantaneous expectation value of the total energy $\langle E_{tot}(t)\rangle $  Eq.~(\ref{eq:exptval}), including the kinetic and potential energies. Open circles: cycle-averaged expectation value of the total energy $\overline{\langle E_{tot}(t)\rangle }$. Each symbol represents the value obtained for an individual optical cycle. For comparison, the pulse intensity envelope $g^2(t)$ is also shown on the respective scales (red thick solid curve). Note that the cycle-averaged total energy gained during the pulse coincides with the ponderomotive potential  $U_p(t)=\mathcal{E}_0^2 g^2(t)/4\omega^2 $  dictated by the pulse envelope as  expected in the high-frequency limit. For instance, the maximal cycle-averaged total energy gained is $ \sim 3.5$~eV at $t = 0$, and  $U_p(0)=3.5 $~eV at the pulse maximum.}\label{fig_ponder3fs}
\end{figure}

In the numerical calculation, we used a Gaussian-shaped pulse of $\tau=3$~fs duration,  $\omega=53.6057$~eV frequency, and peak intensity $I_0=7\times 10^{16}$~W/cm$^2$. At $t=-3\tau=-9$~fs, the initial wave packet is set to the H(1s) ground state function, which is then propagated in the presence of the pulse to $t=3\tau=9$~fs. During the whole propagation time, we have computed the expectation value of the total energy
\begin{equation}
\label{eq:exptval}
\langle E_{tot}(t)\rangle=\langle \Psi(\textbf{r},t) \vert \mathbf{\hat{p}}^2/2+\hat{z}\,\mathcal{E}(t)\vert \Psi(\textbf{r},t) \rangle,
\end{equation}
which includes contributions from both, kinetic energy and potential energy due to interaction with the field. The instantaneous value of $\langle E_{tot}(t)\rangle$ was computed  at 50  time-points for each optical cycle, and then used to obtain the cycle-averaged expectation value  $\overline{\langle E_{tot}(t)\rangle }$.

The  expectation value of the total energy computed numerically as described above for the free electron wave packet is depicted in Fig.~\ref{fig_ponder3fs} as a function of time. The instantaneous value $\langle E_{tot}(t)\rangle$ Eq.~(\ref{eq:exptval}) is shown by thin solid curve in blue. The cycle-averaged expectation value  $\overline{\langle E_{tot}(t)\rangle }$ is depicted in  Fig.~\ref{fig_ponder3fs} by open circles (each circle corresponds to an individual optical cycle).  Clearly, the  cycle-averaged total energy of a free electron increases as the pulse arrives, and it decreases again as the pulse expires. The final total energy after the pulse is off is equal to the initial total energy of the electronic wave packet before the pulse was on. Moreover, the time evolution of the  cycle-averaged total energy follows the pulse intensity envelope $ g^2(t)$, which is also shown in the figure by a thick solid  curve in red to guide the eye.  Importantly, the maximal shift of the cycle-averaged total energy during the pulse is about $\sim 3.5$~eV, and it coincides with the value of $U_p=\mathcal{E}^2_0/4\omega^2\approx 3.5$~eV expected in the high-frequency limit for a  field with an intensity as that at the pulse maximum \cite{B1}. This fact clearly illustrates the accuracy of the present numerical calculations.

The ponderomotive energy in  Fig.~\ref{fig_ponder3fs}  computed for a free electron wave packet (the field-free kinetic energy of 13.6057~eV must be subtracted from the plot) is expected to be larger than that for an electron which is moving in the field of the nucleus because of the attraction it experiences. This is only a qualitative  argument because in the presence of the attractive Coulomb potential, the `ponderomotive motion' of the photoelectron will be a part of the entire photoionization process, and its contribution to the kinetic energy of the emitted photoelectron will not be separable.

\subsection{How fast is the atom ionized by an intense pulse?}
\label{sec:resHOLE}

As a final point, we would like to discuss   which fraction of the electronic wave packet remains bound to the nucleus after the pulse has expired.  As one can see from Fig.~\ref{fig_WP3fs}, a part of the wave packet indeed remains bound to the nucleus (mainly in the ground state of the atom). How to estimate the time-evolution of the population of the ground state? How fast is the hole created during ionization? In order to answer these relevant questions, we turn to our analytical model developed in \cite{DynIntLETT}, which has been proven here to be reliable. As explicitly demonstrated there, the strong pulse induces a time-dependent ionization rate $\Gamma_{ph}(t)$, which describes the losses of the population of the ground state by ionization into all final continuum states. The ionization rate can be computed as the product of the total photoionization cross section at the chosen photon energy $\sigma_{ph}^{tot}(\omega)$ and the photon flux given by $I(t)/\omega$ \cite{RAT1,RAT2}:
\begin{equation}
\label{eq:rate}
\Gamma_{ph}(t)= \sigma_{ph}^{tot}\,I(t)/\omega.
\end{equation}
The rate (\ref{eq:rate}) can further be factorized as $\Gamma_{ph}(t)=\Gamma g^2(t)$. The rate follows the pulse intensity envelope $g^2(t)$ and is the strongest at the pulse maximum. The time-independent parameter explicitly reads $\Gamma=2\pi\vert d_{\omega}\mathcal{E}_0/2\vert^2$ \cite{Demekhin11SFatom}, where $d_\omega$ is the dipole transition matrix element for the ionization of the ground state computed at the chosen photon energy $\omega$.

With the help of this single parameter, one can compute the time evolution of the population of the ground state via the simple analytic expression \cite{DynIntLETT}
\begin{equation}
\label{eq:POP}
\vert a_I(t)\vert^2= e^{-\Gamma \int_{-\infty}^{t}g^2(t^\prime)dt^\prime}.
\end{equation}
This analytic result was found to be in excellent agreement with the results of the full numerical calculations \cite{DynIntLETT}. For the Gaussian-shaped pulses considered here, the final population remaining in the ground state after the pulse has expired can be estimated analytically as $\vert a_I(\infty)\vert^2=e^{-\Gamma \tau  \sqrt{\frac{\pi}{2}}}$. We now can try to introduce the time $T$ needed for the pulse to create the hole as follows. A pulse can ionize only a fraction of all atoms, which is given by $N_h(\infty) =  1-e^{-\Gamma \int_{-\infty}^{\infty}g^2(t)dt} $. When the pulse is over, this fraction defines the hole the pulse has created. At any time $t$, the population of the hole is $N_h(t)=1-e^{-\Gamma \int_{-\infty}^{t}g^2(t^\prime)dt^\prime}$. Following the general concept of a lifetime, we define $T$ as the time at which the not yet populated portion of the hole  $ N_h(\infty)-N_h (T)$ is $1/e$ of the final population of the hole $N_h(\infty)$, i.e., $\left[ N_h(\infty)-N_h (T)\right] / N_h(\infty)  =1/e$. It immediately follows that
\begin{equation}
\label{eq:LTIME}
\int_{-\infty}^{T}g^2(t)dt=\frac{1-\ln \left[ 1+(e-1)\,e^{-\Gamma \int_{-\infty}^{\infty}g^2(t)dt}\right]}{\Gamma},
\end{equation}
from which $T$ is easily computed.

For strong pulses which ionize essentially the whole ensemble of atoms (i.e., $N_h(\infty)  \simeq 1$), condition~(\ref{eq:LTIME}) simplifies
to $\int_{-\infty}^{T}g^2(t)dt \simeq 1/\Gamma$ which is a very appealing result. The simple analytic expressions (\ref{eq:rate}--\ref{eq:LTIME}) allow one to compute the fraction of atoms ionized during the pulse and to estimate the time $T$ needed for the pulse to create the hole. Since the beginning of the pulse is difficult to define, it is convenient to consider this time $T$ relative to the pulse maximum which is in our case at $t=0$. The time $T$ can thus be negative. For the  $\tau=3$~fs Gaussian pulses used in this work, for instance, $T$ is equal to  $+0.27$, $-0.19$, $-0.59$, and  $-0.92$~fs   for the peak intensities $1\times 10^{16}$~W/cm$^2$  to $7\times 10^{16}$~W/cm$^2$ of the spectra
depicted in  Fig.~\ref{fig_ati3fs}, respectively.

\section{Conclusions}
\label{sec:summary}

The time-dependent Schr\"{o}dinger equation for the hydrogen atom exposed to coherent intense high-frequency short laser pulses is solved numerically exactly by directly propagating electron wave packets in space and time. The propagation is made without employing a complex absorption potential
at the spatial grid boundary, and thus requires the use of very large grids even for the presently considered short pulses. In order to solve this technically challenging problem we make use of the efficient code developed in  \cite{DAVID1}, which was additionally optimized for the presently studied problem of an electron in intense pulses. The presently computed electron energy spectra consist of a main photoionization peak and a sequence of above-threshold ionization (ATI) peaks separated by the photon energy $\omega$. Each ATI peak exhibits a pronounced interference pattern which resembles the multiple-peak structure observed in the photoionization peak which is due to dynamic interference \cite{DynIntLETT}.

For the main photoionization peak, the present numerically exact calculations reproduce the results of our previous calculations  \cite{DynIntLETT}  performed by a conceptually very different theoretical approach \cite{Chiang10,Demekhin11SFatom,MolRaSfPRL,DemekhinCO,DemekhinICD,DemekhinBLOCK}. The agreement found makes clear that the assumptions made in the previous calculations are valid. The explicit findings of the present calculations allow us to conclude that the individual AC Stark shifts of the ground state and of the ionization threshold are far from compensating each other, as one would naively assume in the high-frequency strong-field limit \cite{B1}.   The AC Stark effect in the electronic continuum, as well as the dynamic interference effect are found to be rather pronounced.  The present numerical results are analyzed with the help of the analytical model developed in \cite{DynIntLETT}, which allow one to compute the fraction of atoms ionized during the pulse and to estimate the time needed for the pulse to create the hole.

In order to visualize how the free electron wave packet expands in space in short high-frequency pulses and to have a stringent check of the quality of the present numerical calculations, we have also studied the evolution of the wave packet of an electron which is exposed to the pulse, but does not interact with the nucleus.   Our numerical results confirm that in the presence of strong fields, the free electron acquires additional energy which when cycle-averaged can be estimated by the value of the ponderomotive potential $U_p(t)=\mathcal{E}_0^2 g^2(t)/4\omega^2 $  dictated by the pulse envelope and which can be considered as the upper limit for the implicit ponderomotive energy of an electron moving in the field of nucleus. We suggest that in the real system the ponderomotive motion of the photoelectron in the continuum is part of the entire photoionization process, and in strong short pulses its contribution to the energy of the emitted photoelectron is not separable from other effects.

We would like to conclude with the following remark. In the course of the propagation during the pulse, the electron wave packet spreads over large spatial region (see Fig.~\ref{fig_WP3fs}). The parts of the total wave packet emitted on the rising and falling sides of the pulse are always separated in \emph{r}-space and do not meet, as one may naively expect for interference to occur. Nevertheless, these two parts of the wave packet always overlap in \emph{k}-space giving rise to dynamic interference. To measure this interference, one should not perturb by the measurement the evolution of the wave packet in a rather large portion of space.

\begin{acknowledgments}
The authors acknowledge A. Lode for technical support and A.I. Kuleff for many valuable discussions.
\end{acknowledgments}


\begin{thebibliography}{38}
\expandafter\ifx\csname natexlab\endcsname\relax\def\natexlab#1{#1}\fi
\expandafter\ifx\csname bibnamefont\endcsname\relax
  \def\bibnamefont#1{#1}\fi
\expandafter\ifx\csname bibfnamefont\endcsname\relax
  \def\bibfnamefont#1{#1}\fi
\expandafter\ifx\csname citenamefont\endcsname\relax
  \def\citenamefont#1{#1}\fi
\expandafter\ifx\csname url\endcsname\relax
  \def\url#1{\texttt{#1}}\fi
\expandafter\ifx\csname urlprefix\endcsname\relax\def\urlprefix{URL }\fi
\providecommand{\bibinfo}[2]{#2}
\providecommand{\eprint}[2][]{\url{#2}}


\bibitem{KRAUSZ}
F. Krausz and M. Ivanov, Rev. Mod. Phys. \textbf{81}, 163 (2009).

\bibitem{highharm1}
G. Sansone, E. Benedetti, F. Calegari, \emph{et al.},  {Science} \textbf{314}, 443 (2006).

\bibitem{highharm2}
E. Goulielmakis, M. Schultze, M. Hofstetter,  \emph{et. al.},  {Science} \textbf{320}, 1614 (2008).

\bibitem{FLASH}
W. Ackermann,  G. Asova,  V. Ayvazyan, \emph{et al.},  {Nature photonics} \textbf{1}, 336  (2007).

\bibitem{FERMI}
Home page of  FERMI at Elettra in Trieste, Italy, http://www.elettra.trieste.it/FERMI/.

\bibitem{ZEWAIL}
A.H. Zewail, \emph{Femtochemistry, vol. I and II} (World Scientific, Singapore, 1994).

\bibitem{THbp1}
S. Gu\'{e}rin and H. R. Jauslin, Adv. Chem. Phys. \textbf{125}, 147 (2003).

\bibitem{THbp2}
E. Gamaly, \emph{Femtosecond Laser-Matter Interaction: Theory, Experiments and Applications} (Pan Stanford Publishing Pte. Ltd., Singapore, 2011).

\bibitem{THbp3}
B.W. Shore, \emph{Manipulating Quantum Structures Using Laser Pulses} (Cambridge U. P., New York, 2011).

\bibitem{DynIntLETT}
Ph.V. Demekhin   and L.S. Cederbaum, Phys. Rev. Lett. \textbf{108}, 253001 (2012).

\bibitem{Sussman11}
B.J. Sussman,  { Am. J. Phys.} \textbf{79}, 477 (2011).

\bibitem{NOTOURS1}
K. Toyota, O.I. Tolstikhin, T. Morishita,  S. Watanabe, Phys.  Rev. A \textbf{76}, 043418 (2007); \emph{ibid.} \textbf{78}, 033432 (2008).

\bibitem{NOTOURS2}
O.I. Tolstikhin, Phys.  Rev. A \textbf{77},  032712 (2008).

\bibitem{jones}
R.R. Jones, Phys. Rev. Lett. \textbf{74}, 1091 (1995); \emph{ibid.} \textbf{75}, 1491 (1995).

\bibitem{DemekhinDIres}
Ph.V. Demekhin   and  L.S. Cederbaum,  Phys. Rev. A \textbf{86}, 063412 (2012).

\bibitem{Chiang10}
Y.-C. Chiang,  Ph. V. Demekhin, A. I. Kuleff, S. Scheit, and L. S. Cederbaum,  {Phys. Rev. A} \textbf{81}, 032511 (2010).

\bibitem{Demekhin11SFatom}
Ph.V.  Demekhin and L.S. Cederbaum, {Phys. Rev. A } \textbf{83},  023422  (2011).

\bibitem{MolRaSfPRL}
L.S. Cederbaum,  Y.-C. Chiang, Ph.V. Demekhin, and N. Moiseyev,  {Phys. Rev. Lett.} \textbf{106}, 123001 (2011).

\bibitem{DemekhinCO}
Ph.V. Demekhin, Y.-C.  Chiang, and L.S. Cederbaum, {Phys. Rev. A} \textbf{84}, 033417 (2011).

\bibitem{DemekhinICD}
Ph.V. Demekhin, S. D. Stoychev, A.I. Kuleff, and L.S. Cederbaum, Phys. Rev. Lett. \textbf{107}, 273002  (2011).

\bibitem{DemekhinBLOCK}
Ph.V. Demekhin, K. Gokhberg, G. Jabbari, S. Kopelke, A.I. Kuleff, and L.S. Cederbaum,  J. Phys. B  \textbf{46}, 021001  (2013).

\bibitem{POPOV1}
E.A. Volkova, V.V. Gridchin, A.M. Popov, and O.V. Tikhonova, JETP  \textbf{102}, 40 (2006).

\bibitem{POPOV2}
E.A. Volkova, A.M. Popov, M.A. Tikhonov, and O.V. Tikhonova, JETP \textbf{105}, 526 (2007).

\bibitem{POPOV3}
A.M. Popov, M.A. Tikhonov, and O.V. Tikhonova, and E.A. Volkova, Laser Physics  \textbf{19}, 191 (2009).

\bibitem{C3el1}
J. Colgan and M. S. Pindzola, Phys. Rev. Lett. \textbf{108}, 053001 (2012).

\bibitem{C3el2}
J. Colgan, A. Emmanouilidou, and M.S. Pindzola, Phys. Rev. Lett. \textbf{110}, 063001 (2013).

\bibitem{C2el}
M.S. Pindzola, C.P. Ballance, Sh.A. Abdel-Naby, F. Robicheaux, G.S.J. Armstrong, and J. Colgan, J. Phys. B \textbf{46}, 035201 (2013).

\bibitem{Clgr}
F. Robicheaux, J. Phys. B \textbf{45}, 135007 (2012).

\bibitem{MCTDH}
H.-D. Meyer, U. Manthe, and L.S. Cederbaum, Chem. Phys. Lett. \textbf{165}, 73 (1990).

\bibitem{MCTDH1}
U. Manthe, H.-D. Meyer, and L.S. Cederbaum, J. Chem. Phys.  \textbf{97}, 3199 (1992); \emph{ibid.}  \textbf{97}, 9062 (1992).

\bibitem{MCTDHBOOK}
H.-D. Meyer, F. Gatti, and G. A. Worth (Eds.)\emph{ Multidimensional Quantum Dynamics: MCTDH Theory and Applications} (Wiley-VCH, Weinheim, 2009).

\bibitem{BOSON1}
O.E. Alon, A.I. Streltsov, and L.S. Cederbaum, Phys. Rev. A 77, 33613
(2008).

\bibitem{FERM1}
J. Zanghellini, M. Kitzler, C. Fabian, T. Brabec, and A. Scrinzi, Laser Phys. \textbf{13}, 1064 (2003).

\bibitem{FERM2}
T. Kato and H. Kono, Chem. Phys. Lett. \textbf{392}, 533 (2004).

\bibitem{FERM3}
M. Nest, T. Klamroth, and P. Saalfrank, J. Chem. Phys. \textbf{122}, 24102 (2005).

\bibitem{FERM4}
D.J. Haxton, K.V. Lawler, and C.W. McCurdy, Phys. Rev. A \textbf{83}, 063416 (2011).

\bibitem{DAVID1}
D. Hochstuhl and M. Bonitz, J. Chem. Phys. \textbf{134}, 084106 (2011).

\bibitem{BOSON2}
O. E. Alon, A. I. Streltsov, and L. S. Cederbaum, J. Chem. Phys. \textbf{127}, 54103 (2007).

\bibitem{PR00}
M.H. Beck, A. J\"{a}ckle, G.A. Worth, and H.D. Meyer, Phys. Rep. \textbf{324}, 1 (2000).

\bibitem{FEDVR}
T.N. Rescigno and C.W. McCurdy, Phys. Rev. A \textbf{62}, 032706 (2000).

\bibitem{ALG1}
T.J. Park and J.C. Light, J. Chem. Phys. \textbf{85}, 5870 (1986).

\bibitem{B1}
M.V. Fedorov,  \emph{Atomic and free electrons in a strong light field} (World Scientific, River Edge, 1997).

\bibitem{B2}
K.C. Kulander, M. Lewenstein, \emph{`Multiphoton and Strong-Field Processes' in Handbook of Atomic, Molecular, and Optical Physics}, ed. by G.W.F. Drake (Springer, Heidelberg, 2006).

\bibitem{RAT1}
Y.-P. Sun, J.-C. Liu,  C.-K. Wang, and F. K. Gel'mukhanov F K Phys. Rev. A \textbf{81}, 013812 (2010).

\bibitem{RAT2}
J.-C. Liu, Y.-P. Sun, C.-K. Wang, H. \AA gren, and F. K. Gel'mukhanov, Phys. Rev. A \textbf{81}, 043412 (2010).

\end{thebibliography}
\end{document}